\documentclass[twocolumn,prl,showpacs]{revtex4}%
\usepackage{graphicx}%
\usepackage{amsmath}%
\usepackage{amsfonts}%
\usepackage{amssymb}

\def\s{{\sigma}}

\def\k{{ {\bf k} }}

\def\l{{\lambda}}

\begin{document}
\title{ Giant Intrinsic Spin and Orbital Hall Effects in Sr$_{2}M$O$_{4}$
($M$=Ru,Rh,Mo) }
\author{H. \textsc{Kontani}$^{1}$, T. \textsc{Tanaka}$^{1}$, D.S. \textsc{Hirashima}%
$^{1}$, K. \textsc{Yamada}$^{2}$ and J. \textsc{Inoue}$^{3}$ }
\date{\today }

\begin{abstract}
We investigate the intrinsic spin Hall conductivity (SHC) and the $d$-orbital Hall
conductivity (OHC) in metallic $d$-electron systems, by focusing on the
$t_{2g}$-orbital tight-binding model for Sr$_{2}M$O$_{4}$ ($M$=Ru,Rh,Mo). The
conductivities obtained are one or two orders of magnitude larger than
predicted values for $p$-type semiconductors with $\sim5$\% hole doping. The
origin of these giant Hall effects is the ``effective Aharonov-Bohm phase''
that is induced by the $d$-atomic angular momentum in connection with 
the spin-orbit interaction and the inter-orbital hopping integrals.
The huge SHC and OHC generated by this mechanism are
expected to be ubiquitous in multiorbital transition metal complexes, which
opens the possibility of realizing spintronics as well as
\textquotedblleft orbitronics\textquotedblright\ devices.

\end{abstract}

\maketitle


\address{
$^1$Department of Physics, Nagoya University,
Furo-cho, Nagoya 464-8602, Japan. \\
$^2$Engineering, Ritsumeikan University,
1-1-1 Noji Higashi, Kusatsu, Shiga 525-8577, Japan. \\
$^3$Department of Applied Physics, Nagoya University,
Furo-cho, Nagoya 464-8602, Japan.
}

\sloppy


Recently, the spin Hall effect (SHE) has been attracting considerable 
interest since controlling spin by applying electric field
is one of the key issues in the field of spintronics \cite{Loss}. 
The SHE was originally predicted based on the assumption of extrinsic 
electron-scattering \cite{dyakonov,hirsh}.
Intrinsic SHE in semiconductors is caused by uniform spin-orbit interaction 
(SOI) in systems and is independent of the presence of impurities.
It existence was predicted by Murakami \textit{et al}.\cite{Murakami} 
for $p$-type semiconductors, and by Sinova \textit{et al}.\cite{Niu04} 
for a two-dimensional electron gas (2DEG) with a Rashba-type SOI. 
Motivated by these predictions, many studies have been
performed to demonstrate the prominent disorder effects on the
spin Hall conductivity (SHC) \cite{inoue1,murakami2,bernevig2,inoue2}, 
the possibility of quantum SHE in insulating phase of semiconductors
\cite{murakami3,kane06,bernevig06}, 
and the existence of the orbital Hall effect (OHE).
According to the result of the \textit{ab initio} calculation
\cite{Niu05}, SHC and orbital Hall conductivity (OHC) for $p$-type
semiconductors increase with hole number $n_{\mathrm{h}}$, and they reach
about 100 and 10 [$\hbar e^{-1} \cdot\Omega^{-1}$cm$^{-1}$] respectively, for
$n_{\mathrm{h}}=0.05/\mathrm{cell}$.

Very recently, the so-called inverse SHE has been observed for spin currents
injected into conventional metals such as Al and Pt from a ferromagnet
\cite{valenzuela,kimura}. The SHC observed in Pt is $\sim240\ [\hbar e^{-1}
\cdot\Omega^{-1}\mathrm{cm}^{-1}]$, which is about $10^{4}$ times larger than
the experimental SHC in $n$-type semiconductors reported in ref.
\cite{wunderlich}. This observation raises the question of whether a
novel mechanism may be responsible for the \textit{giant SHE} in $d$-electron
systems. Moreover, a simple theoretical model that is capable of describing 
the giant SHE inherent in $d$-electron systems is highly desirable.

\begin{figure}[ptbh]
\begin{center}
\includegraphics[width=.7\linewidth]{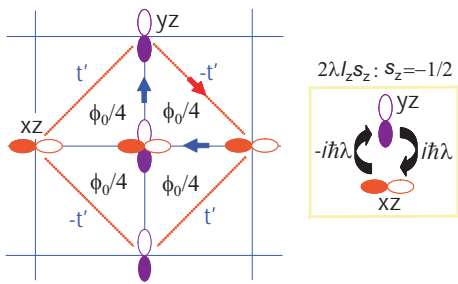}
\end{center}
\caption{ ``Effective magnetic flux'' for $\downarrow$-electron through
triangles of half unit cells in the $(d_{xz},d_{yz})$-model. The effective
magnetic flux for $\uparrow$-electron is $-\phi_{0}/4$. }%
\label{fig:flux}%
\end{figure}

In this Letter, we propose a new mechanism for the giant intrinsic SHC and OHC
that originate from the $d$-orbital degrees of freedom, which is absent in
semiconductors. In a multiorbital system, a conduction electron acquires an
\textquotedblleft effective Aharonov-Bohm phase factor\textquotedblright\ due
to $d$-atomic angular momentum in conjunction with 
the SOI and the inter-orbital hoppings.
This  mechanism is responsible for the large SHE.
An intuitive explanation based on the tight-binding model,
which is given in Fig. \ref{fig:flux}, appears to successfully predict 
the characteristics of SHE in $d$-electron systems. 
We stress that not only the SOI play a significant role in the SHE, but so does
the phase of hopping integrals characteristic of $d$-electron systems.
In contrast, the \textquotedblleft Dirac
monopole mechanism\textquotedblright\ \cite{Murakami} is appropriate when
massless Dirac cone dispersion exists in close proximity to the Fermi level,
as is the case in semiconductors. 
The present mechanism of SHE also differs from that in the Rashba-type 
2DEG model due to the momentum-dependent SOI \cite{Niu04}. We further
show that the calculated SHC depends strongly on the longitudinal resistivity
in the high-resistivity region, even though the SHE is of intrinsic origin.

To demonstrate the mechanism of the intrinsic SHC and OHC 
due to $d$-orbital degrees of
freedom, we adopt the square-lattice tight-binding model with ($d_{xz}%
,d_{yz},d_{xy}$)-orbitals, which was first proposed to describe the realistic
electron-structure of Sr$_{2}M$O$_{4}$ ($M$=Ru,Rh,Mo) \cite{Sigrist}, and
atomic LS coupling as the SOI. 
This model is subsequently referred to as the $t_{2g}$-model.
We show that the calculated SHC and OHC are one or two orders of magnitude
greater than theoretical values in highly doped $p$-type semiconductors
\cite{Niu05}.
The present study strongly predicts that giant SHE and OHE are
ubiquitous in $d$-electron systems, and suggests a new method to analyze 
them in various metals.

The Hamiltonian for the $t_{2g}$-model in a square lattice
is given by $H=H_{\mathrm{K}}+H_{\mathrm{SO}}$, where $H_{\mathrm{K}}$ 
is the kinetic term and
$H_{\mathrm{SO}}=\sum_{\imath}2{\lambda}{{{\mbox{\boldmath $l$}}}}_{\imath
}\!\cdot\!{{{\mbox{\boldmath $s$}}}}_{i}$ is the SOI for $M$ $4d$-orbitals. The
band parameters of the $t_{2g}$-model are: (i) the intraorbital nearest neighbor
(NN) hopping integrals $t_{1}=t_{2}(\equiv t)$ and $t_{3}$ for the $xz,yz$ and
$xy$ orbitals, respectively, (ii) intraorbital second NN hopping $t_{3}^{\prime}$
for $xy$ orbital, and (iii) interorbital second NN hopping $t^{\prime}$ between
the $xz$ and $yz$ orbitals. 
It will be shown later that interorbital hopping
$t^{\prime}$\ is indispensable for SHE and OHE.

The matrix elements of $H_{K}$ are then easily given by
\begin{equation}
{\hat{H}}_{K}=\left(
\begin{array}
[c]{ccc}%
\xi_{1} & g & 0\\
g & \xi_{2} & 0\\
0 & 0 & \xi_{3}%
\end{array}
\right)  ,
\end{equation}
where $\xi_{1}=-2t\cos k_{x}$, $\xi_{2}=-2t\cos k_{y}$, $\xi_{3}=-2t_{3}(\cos
k_{x}+\cos k_{y})-4t_{3}^{\prime}\cos k_{x}\cos k_{y}+\xi_{3}^{0}$, and
$g=-4t^{\prime}\sin k_{x}\sin k_{y}$. Here, the first, second, and 
third rows (columns) of ${\hat{H}}_{K}$ correspond to $|xz\rangle$, $|yz\rangle
$, and $|xy\rangle$, respectively, and 1, 2, and 3 denote orbitals $xz$, $yz$, and
$xy$, respectively. 
We set $\hbar=1$ during the calculation. 
A constant $\xi_{3}^{0}$ is included in $\xi_{3}$ 
to allow the number of electrons $n_{l}$ in the $l$-orbital to be adjusted.

\begin{figure}[ptbh]
\includegraphics[width=.7\linewidth]{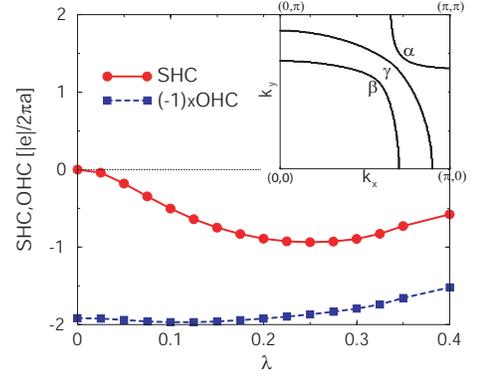}\caption{ ${\lambda}$-dependence of
the SHC and OHC in Sr$_{2}$RuO$_{4}$. A typical value of ${\lambda}$ for
Ru$^{4+}$-ion is $\sim0.2$.  We set
${\gamma}_{1}={\gamma}_{2}=0.005$ and ${\gamma}_{3}/{\gamma}_{1}=3$. 
Inset: Fermi surfaces of Sr$_{2}$RuO$_{4}$ for $\lambda=0.2$.
}%
\label{fig:LS}%
\end{figure}

The total Hamiltonian is given by
\begin{equation}
{\hat{H}}=\left(
\begin{array}
[c]{cc}%
{\hat{H}}_{K}+{\lambda}{\hat{l}}_{z} & {\lambda}({\hat{l}}_{x}-i{\hat{l}}%
_{y})\\
{\lambda}({\hat{l}}_{x}+i{\hat{l}}_{y}) & {\hat{H}}_{K}-{\lambda}{\hat{l}}_{z}%
\end{array}
\right)  ,
\end{equation}
where the first and the second rows (columns) correspond to $|\!\uparrow\rangle$
and $|\!\downarrow\rangle$, respectively.
The matrix element of ${\hat{l}}_{\zeta}$ ($\zeta=x,y,z$) is specified in
quantum mechanics textbooks as, for example, 
$({\hat{l}}_{z})_{l,m}=i(\delta_{l,2}\delta_{m,1}-\delta_{l,1}\delta_{m,2})$. 
In accordance with Refs.
\cite{Sigrist,Nomura,Mizokawa}, 
we set $t=1$ ($\approx0.2$ eV),
$t^{\prime}=0.1$, $t_{3}=0.8$, $t_{3}^{\prime}=0.35$, and ${\lambda}\sim0.2$.
The band structure for Sr$_{2}$RuO$_{4}$ is reproduced by selecting the
chemical potential $\mu$ and $\xi_{3}^{0}$ so that $n_{1}=n_{2}=n_{3}=n/3$. 
The Fermi surfaces for ${\lambda}=0.2$ are shown in the inset of
Fig. \ref{fig:LS}. (${\alpha}$,${\beta}$)-bands [${\gamma}$-band] are mainly
composed of ($d_{xz},d_{yz}$)-orbitals [$d_{xy}$-orbital]
and, due to the symmetry of ${\hat H}$ under $\k\leftrightarrow-\k$,
each band is two-fold degenerate
even if ${\lambda}>0$, in contrast to the Rashba 2DEG model. 
The shape of the Fermi surfaces agrees with that obtained in the ARPES
measurements, LDA band calculations, and dHvA oscillations \cite{Maeno}.

The $6\times6$ matrix retarded Green function is given by ${\hat{G}}%
^{R}({{\ \mathbf{k}}},{\omega})=({\omega}+\mu-{\hat{H}}+i{\hat{\Gamma}})^{-1}%
$, where ${\hat{\Gamma}}$ is the imaginary part of the 
${\ \mathbf{k}}$-independent self-energy due to local impurities. 
Since a tiny residual resistivity in Sr$_{2}$RuO$_{4}$ suggests that the 
impurity potentials are small, we apply the Born approximation:
${\hat{\Gamma}}\propto-\mathrm{Im}\sum_\k {\hat G}^{R}(\k,0)$. 
If ${\lambda}=0$, then ${\hat{\Gamma}}$ is diagonal; 
${\hat{\Gamma}}_{l,m}\propto\gamma_{l}\delta_{l,m}$ ($l,m=1,2,3$). 
The offdiagonal terms will then be negligible if
${\lambda}\ll W_{\mathrm{band}}$. ${\gamma}_{l}$ is the quasiparticle damping
rate for the $l$-orbital, and is proportional to the local density of states
(LDOS) for the $l$-orbital,
$\rho_{l}(0)$. In Sr$_{2}$RuO$_{4}$, $1\ll\rho_{3}(0)/\rho_{1}(0)\sim2.5$ 
since the $\gamma$-band Fermi surface is very close to the
van-Hove singular point at $(\pi,0)$ \cite{Sigrist}.

The charge current for $\zeta$-direction ($\zeta=x,y$) is given by ${\hat{J}%
}_{\zeta}^{\mathrm{C}}=-e\cdot d{\hat{H}}/dk_{\zeta}$ ($-e$ is the electron
charge), and the ${\sigma}_{z}$-spin current is given by the 
Hermitian operator:
${\hat{J}}_{\zeta}^{\mathrm{S}}
=(-1/e)\{{\hat{J}}_{\zeta}^{\mathrm{C}},{\hat{s}}_{z}\}/2$
 \cite{Murakami,Niu04,inoue1,murakami2,bernevig2,inoue2,murakami3,kane06,bernevig06,Niu05}.
Since the SOI is ${\ \mathbf{k}}$-independent, they are given by
\begin{equation}
{\hat{J}}_{\zeta}^{\mathrm{C}}=-e\left(
\begin{array}
[c]{cc}%
{\hat{v}}_{\zeta} & 0\\
0 & {\hat{v}}_{\zeta}%
\end{array}
\right)  ,\ \ {\hat{J}}_{\zeta}^{\mathrm{S}}=\frac{1}{2}\left(
\begin{array}
[c]{cc}%
{\hat{v}}_{\zeta} & 0\\
0 & -{\hat{v}}_{\zeta}%
\end{array}
\right)  ,
\label{eqn:J}
\end{equation}
where ${\hat{v}}_{\zeta}=d{\hat{H}}_{K}/dk_{\zeta}$. Also, the $l_{z}$-orbital
current is given by ${\hat{J}}_{\zeta}^{\mathrm{O}}=\{{\hat{J}}_{\zeta}^{\mathrm{C}},{\hat{l}}_{z}\}/2$
 \cite{Niu05}.

\begin{figure}[ptbh]
\includegraphics[width=.42\linewidth]{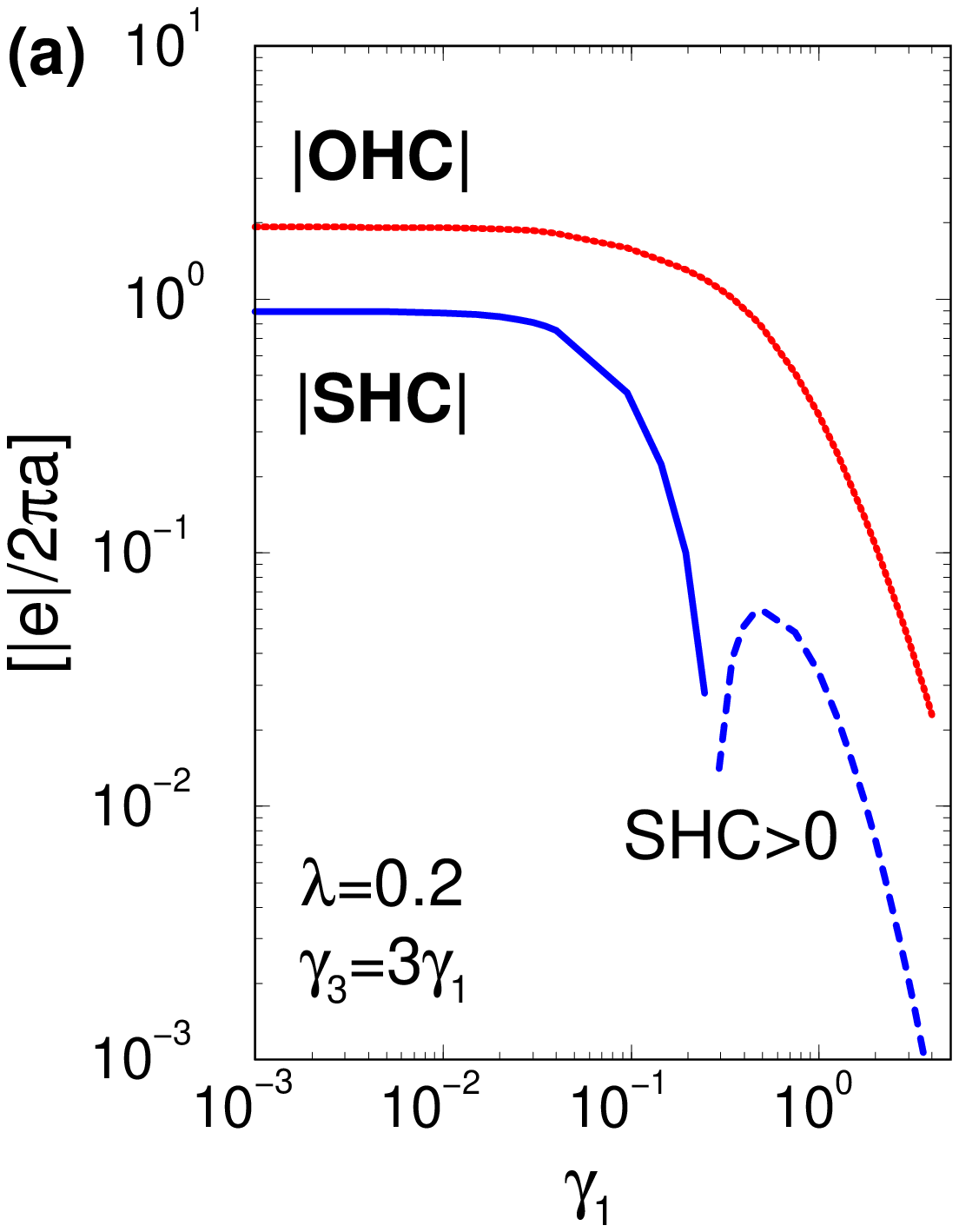}
\includegraphics[width=.42\linewidth]{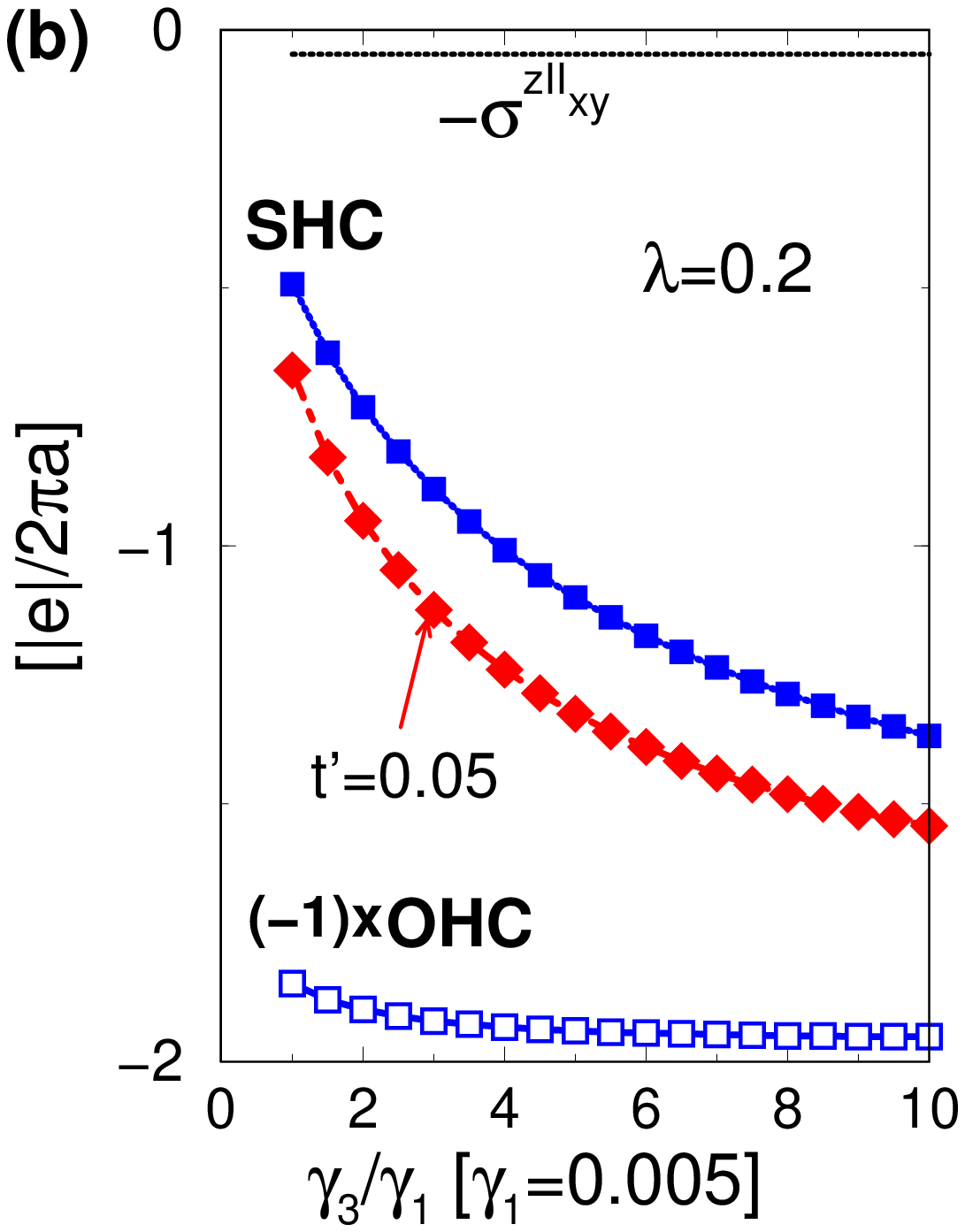}
\includegraphics[width=.60\linewidth]{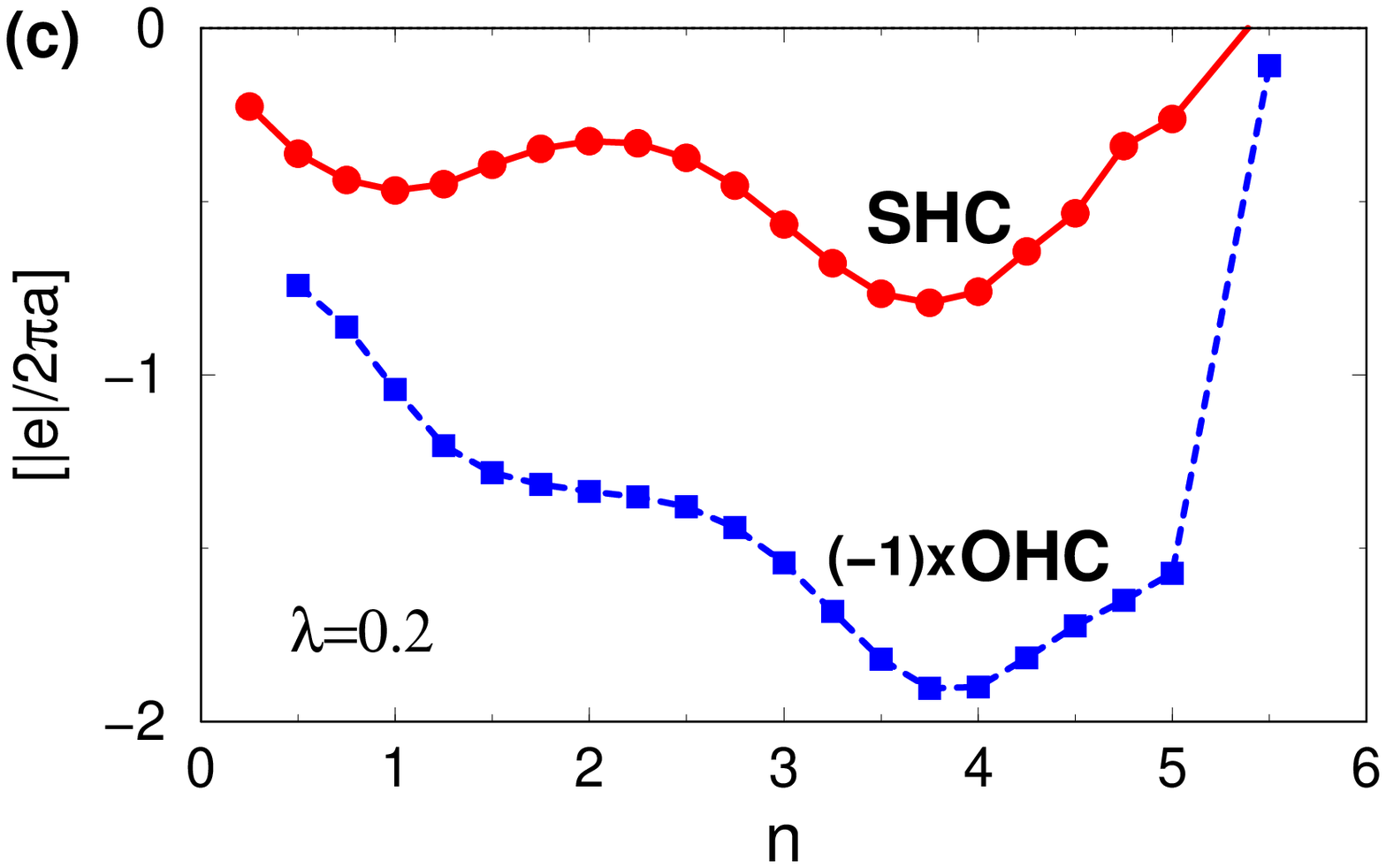}\caption{ 
(a) ${\gamma}_{1}$-dependence and 
(b) ${\gamma}_{3}/{\gamma}_{1}$-dependence of the $|$SHC$|$ and $|$OHC$|$
in Sr$_{2}$RuO$_{4}$. $|{\sigma}_{xy}^{z}(t^{\prime}=0.05)|$ is larger than
$|{\sigma}_{xy}^{z}(t^{\prime}=0.1)|$, which is understood from the relation
$|{\sigma}_{xy}^{z}| \propto t^{\prime}\Delta^{-2}$ in the present model. (c)
$n$-dependence of the SHC and OHC in the $t_{2g}$-model [Sr$_{2}M$O$_{4}$]. }
\label{fig:Gratio}
\end{figure}

We now calculate the SHC and OHC in the presence of local impurities using
the linear-response theory. 
In the current model, ${\hat{v}}_{\zeta}$ and $G $
are respectively odd and even functions with respect to ${\ \mathbf{k}%
}\leftrightarrow-{\ \mathbf{k}}$. Therefore, the current vertex correction
(CVC) due to the local impurity potentials in the (extended) Born
approximation, which is given by 
$\Delta{\hat{J}}_{\zeta}^{\mathrm{M}}
\propto \sum_{\k} {\hat{G}}^{A} {\hat{J}}_{\zeta}^{\mathrm{M}}{\hat{G}}^{R}$
(M=S,C or O), vanishes identically.
(In contrast, the CVC is crucial in the Rashba 2DEG model since the SOI, which is
an odd-function with respect to ${\ \mathbf{k}}$, gives a ${\ \mathbf{k}}%
$-independent anomalous velocity \cite{inoue1}.)
For this reason, the SHC at
$T=0$ is given by ${\sigma}_{xy}^{z}={\sigma}_{xy}^{zI}+{\sigma}_{xy}^{zI\!I}$
\cite{Streda}, where
\begin{align}
{\sigma}_{xy}^{zI}  & =\frac{1}{2\pi N}\sum_{{\ \mathbf{k}}}\mathrm{Tr}\left[
{\hat{J}}_{x}^{\mathrm{S}}{\hat{G}}^{R}{\hat{J}}_{y}^{\mathrm{C}}{\hat{G}}%
^{A}\right]  _{{\omega}=0},\label{eqn:SHCI}\\
{\sigma}_{xy}^{zI\!I}  & =\frac{-1}{4\pi N}\sum_{{\ \mathbf{k}}}\int_{-\infty
}^{0}d{\omega}\mathrm{Tr}\left[  {\hat{J}}_{x}^{\mathrm{S}}\frac{{\hat{G}}%
^{R}}{{\partial}{\omega}}{\hat{J}}_{y}^{\mathrm{C}}{\hat{G}}^{R}\right.
\nonumber\\
& \left.  \ \ \ \ \ \ \ \ -{\hat{J}}_{x}^{\mathrm{S}}{\hat{G}}^{R}{\hat{J}%
}_{y}^{\mathrm{C}}\frac{{\hat{G}}^{R}}{{\partial}{\omega}}-\langle
\mathrm{R}\rightarrow\mathrm{A}\rangle\right]  .\label{eqn:SHCII}%
\end{align}
Here, $I$ and $I\!I$ represent the \textquotedblleft Fermi surface
term\textquotedblright\ and the \textquotedblleft Fermi sea
term\textquotedblright. 
Taking the trace in eq. (\ref{eqn:SHCI}), ${\sigma
}_{xy}^{zI}$ in the present model is straightforwardly given by
${\sigma}_{xy}^{zI(1)}+{\sigma}_{xy}^{zI(2)}+{\sigma}_{xy}^{zI(3)}$, where
\begin{eqnarray}
\s_{xy}^{zI(1)}&=& \frac{2e\l}{\pi N}\sum_\k \gamma_1 v_x^a v_y 
 ((\mu-\xi_3)^2+\gamma_3^2)/|D^R|^2,
 \label{eqn:Sxy-r1} \\
\s_{xy}^{zI(2)}&=& \frac{e\l^2}{\pi N}\sum_\k 
 \left[ 2\gamma_1 v_x^a v_y (\mu-\xi_3) \right.
 \nonumber \\
& & + \left. \gamma_3 (g v_x v_y +2v_x^a v_y (\mu-\xi_3)) 
 \right]/|D^R|^2,
 \label{eqn:Sxy-r2} \\
\s_{xy}^{zI(3)}&=& \frac{2e\l^3}{\pi N}\sum_\k \gamma_3 v_x^a v_y/
 |D^R|^2,
 \label{eqn:Sxy-r3}
\end{eqnarray}
where $v_{\zeta}^{a}=dg/dk_{\zeta}$ ($\zeta=x,y$), $v_{x}=d\xi_{1}/dk_{x} $,
$v_{y}=d\xi_{2}/dk_{y}$, and $D^{R}=\mathrm{det}(\mu-{\hat{H}}^{s}%
+i{\hat{\Gamma}})$.
${\sigma}_{xy}^{zI(m)}$ ($m=1,2,3$) is proportional to ${\lambda}^{m}$ except
for the ${\lambda}$-dependence of $D^{R}$. The \textquotedblleft Fermi sea
term\textquotedblright\ can also be obtained by inserting 
${\hat{J}}_{x}^{\mathrm{C,S}}$ in eq. (\ref{eqn:J}) into eq. (\ref{eqn:SHCII}).
We do not give an expression for it since its contribution is very small in the 
metallic state \cite{Kontani06} (see Fig. \ref{fig:Gratio} (b)).

In the same way, the OHC of the Fermi surface term $O_{xy}^{zI}$ and that of
the Fermi sea term $O_{xy}^{zI\!I}$ are respectively given by eqs.
(\ref{eqn:SHCI}) and (\ref{eqn:SHCII}) by replacing ${\hat J}_{x}^{\mathrm{S}%
}$ with ${\hat J}_{x}^{\mathrm{O}}$.
In the present model, $O_{xy}^{zI}$ is given by $O_{xy}^{zI(0)}+O_{xy}%
^{zI(2)}$, where
\begin{eqnarray}
\!\!\!\!\!\! O_{xy}^{zI(0)}\!&=& \frac{e}{\pi N}\sum_\k \gamma_1
 v_x \left[g v_y + v_y^a (\xi_1-\xi_2) \right] 
 \label{eqn:Oxy-r0} \\
& &\times ((\mu-\xi_3)^2+\gamma_3^2)/|D^R|^2,
 \nonumber \\
\!\!\!\!\!\! O_{xy}^{zI(2)}\!&=& \frac{e\l^2}{\pi N}\sum_\k \gamma_3 
 v_x \left[g v_y + v_y^a (\xi_1-\xi_2) \right]/|D^R|^2.
 \label{eqn:Oxy-r2}
\end{eqnarray}

The interorbital velocity $v_{\zeta}^{a}=dg/dk_{\zeta}$ is termed the
\textquotedblleft anomalous velocity\textquotedblright, which is the origin of
the Hall effect \cite{Kontani06}. Since $v_{x}^{a}=-4t^{\prime}\sin k_{y}\cos
k_{x}$ has the same symmetry as $k_{y}$, eqs. (\ref{eqn:Sxy-r1}%
)-(\ref{eqn:Sxy-r3}) are finite even after the ${\ \mathbf{k}}$-summations.
Since $t^{\prime}$ is much larger than the Rashba SOI in semiconductors,
a large SHC is realized in the present model. Although the sign of $t^{\prime}$
changes due to the gauge transformation $|xz\rangle\rightarrow-|xz\rangle$,
both the SHC and OHC are invariant
if the sign change of $({\hat{l}}_{\zeta})_{l,1}$ is taken into account
\cite{Kontani06}. We stress that eqs. (\ref{eqn:Sxy-r1})-(\ref{eqn:Oxy-r2})
are independent of the renormalization factor due to the Coulomb interaction:
$z=(1-{\partial}\Sigma({\omega})/{\partial}{\omega})^{-1}|_{{\omega}=0}$
\cite{Kontani06}.

Figure \ref{fig:LS} shows the ${\lambda}$-dependence of the SHC and OHC for
the $t_{2g}$-model with $M$=Ru ($n=4$). Here, we put 
${\gamma}_{1}={\gamma}_{2}=0.005$ and ${\gamma}_{3}/{\gamma}_{1}=3$. 
The Fermi sea terms ($I\!I$) of
both the SHC and OHC are negligible \cite{Kontani06} since they are about 30
times smaller than the Fermi surface terms ($I$) given in eqs.
(\ref{eqn:Sxy-r1})-(\ref{eqn:Oxy-r2}) (see Fig. \ref{fig:Gratio} (b)). 
Here, $1.0\ [e/2\pi a]$ corresponds to $\approx670\ [\hbar e^{-1}\cdot\Omega
^{-1}\mathrm{cm}^{-1}]$ if we put the interlayer distance of Sr$_{2}M$O$_{4}$,
$a\approx6$\AA. The obtained SHC and OHC for typical values of ${\lambda
}\sim0.2$ are much larger than those in semiconductors \cite{Niu05,Murakami}.
In Fig. \ref{fig:LS}, the SHC increases approximately in proportion to
${\lambda}^{2}$ since ${\sigma}_{xy}^{zI(2)}\ $in eq. (\ref{eqn:Sxy-r2}) gives
the dominant contribution for $0.05<{\lambda}<0.3$. On the other hand, the OHC
exists even if ${\lambda}=0$ since the orbital current is independent of spin.
Note that the OHC vanishes if ${\lambda}=0$ in Rashba 2DEG model since
the anomalous velocity arises from the ${\ \mathbf{k}}$-dependent SOI.

Figure \ref{fig:Gratio} (a) shows the ${\gamma}_{1}$-dependences of the SHC
and OHC for ${\gamma}_{3}/{\gamma}_{1}=3$.
They are approximately independent
of ${\gamma}_{1}$ in the low-resistivity region where ${\gamma}_{1}<0.1$. 
This behavior is typical of the intrinsic Hall effect found by Karplus and
Luttinger \cite{KL}. 
However, both the SHC and OHC start to decrease with
${\gamma}_{1}$ in the high-resistivity region where ${\gamma}_{1}>0.1$. This
crossover behavior is also realized in the intrinsic anomalous Hall effect
(AHE) \cite{Kontani94,Kontani06}:
Since the intrinsic Hall effect originates from the interband 
particle-hole excitation induced by the electric field \cite{KL}, 
the Hall conductivity is proportional to the lifetime of the
excitation $\hbar/\Delta$, where $\Delta$ is the minimum band-splitting around
the Fermi level \cite{Kontani06}. 
In this model, $\Delta\sim0.2$.
In the high-resistivity region, the SHC decreases drastically since the
interband excitation is suppressed when the quasiparticle lifetime
$\hbar/{\gamma}$ is shorter than $\hbar/\Delta$. 
Interestingly, the SHC becomes
positive for ${\gamma}_{1}\ge0.3$ since ${\sigma}_{xy}^{zI(1)} \ (>0)$ is
dominant over ${\sigma}_{xy}^{zI(2)}$.

Even in the low-resistivity region, the SHC and OHC sensitively depend on
the value of ${\gamma}_{3}/{\gamma}_{1}$ as shown in Fig. \ref{fig:Gratio}
(b). $|{\sigma}_{xy}^{z}|$ increases significantly with ${\gamma}_{3}/{\gamma
}_{1}$ since the numerator of the second term in eq. (\ref{eqn:Sxy-r2}) and
that in eq. (\ref{eqn:Sxy-r3}) are proportional to ${\gamma}_{3}$.
On the other hand, $O_{xy}^{z}$ is approximately independent of ${\gamma}%
_{3}/{\gamma}_{1}$. Here, we consider how to increase ${\gamma}_{3}/{\gamma
}_{1}$ in real systems. 
According to ref. \cite{La}, $\gamma$-band Fermi
surface touches $(\pi,0)$ in La-doped Sr$_{2-y}$La$_{y}$RuO$_{4}$ for
$y\sim0.23$ ($n\sim4.23$). Then, ${\gamma}_{3}/{\gamma}_{1}\gg1$ is expected
within the Born approximation. We also consider Ca-doped Ca$_{2-y}$Sr$_{y}%
$RuO$_{4}$ ($n=4$):
According to optical measurements \cite{Ca}, a huge mass-enhancement is
realized in $\gamma$-band, whereas ${\alpha},{\beta}$-bands remain light. In
this situation, ${\gamma}_{3}/{\gamma}_{1}\gg1$ due to inelastic scattering 
is expected at finite temperatures.

Figure \ref{fig:Gratio} (c) shows the electron filling ($n$) dependence of the
SHC and OHC in the $t_{2g}$-model. For Sr$_{2}M$O$_{4}$, $M$=Ru, Mo and Rh
correspond to $n=4$, 2 and 5, respectively \cite{Mo,Rh}. 
Here, we correctly account for the $n$-dependence of 
${\gamma}_{3}/{\gamma}_{1} = \rho_{3}(0)/\rho_{1}(0)$.
It is thus possible to control the values of SHC and OHC by varying the
electron filling.

We now present an intuitive explanation for the SHE in $d$-electron systems
based on the $(d_{xz},d_{yz})$-model \cite{Miyazawa,Kontani06}, which is a
simplified version of the present model. Here, all the matrix elements of
${\hat{l}}_{x}$ and ${\hat{l}}_{y}$ vanish identically. In Fig. \ref{fig:flux},
we consider the motion of a $\downarrow$-spin electron along a triangle of
a half unit cell. An electron in the $d_{xz}$-orbital can transfer to the 
$d_{yz}$-orbital and vice versa using the SOI for $\downarrow$-electron
$-\hbar{\lambda}{\hat{l}}_{z}$, $\langle yz|{\hat{l}}_{z}|xz\rangle=-\langle
xz|{\hat{l}}_{z}|yz\rangle=i$. These relations are derived from the fact that
the $(\pi/2)$-rotation operator about the $z$-axis is given by $\exp
(-i\pi{\hat{l}}_{z}/2)=\cos(\pi/2)-i{\hat{l}}_{z}\sin(\pi/2)=-i{\hat{l}}_{z}$
for ${\hat{l}}_{z}^{2}=1$. By considering the sign of the interorbital hopping
integral ($\pm t^{\prime}$) and the matrix elements of the SOI, 
it can be shown that a clockwise (anti-clockwise) motion along any triangle 
of the half unit cell causes the factor $+i$ ($-i$).

This factor can be interpreted as the Aharonov-Bohm phase factor $e^{2\pi
i\phi/\phi_{0}}$ [$\phi_{0}=hc/e$], where $\phi= \oint\mathbf{A}
d\mathbf{r}=\phi_{0}/4$ represents the ``effective magnetic flux'' through the
triangle of the half unit cell.
Since the effective magnetic flux for $\uparrow$-electron is opposite in sign,
$\uparrow$-electrons and $\downarrow$-electrons move in opposite directions. 
Thus, the effective magnetic flux in Fig. \ref{fig:flux} gives rise
to a SHC of order $O(\lambda)$ [eq. (\ref{eqn:Sxy-r1})], which is analogous
to the AHE due to spin chirality \cite{Taillefumier,Ohgushi}. 
(In the $t_{2g}$-model, the SHC of order $O(\lambda^{2})$ in 
eq. (\ref{eqn:Sxy-r2}) is derived from the interorbital transition between 
($xy\!\uparrow$) and ($xz\!\downarrow$) [or ($yz\!\downarrow$)] 
using the SOI twice.) 
We can also discuss the origin of the OHE by considering the motion of 
an electron with $|l_{z}=\pm1\rangle\propto\mp|xz \rangle+ i |yz \rangle$ 
in the absence of the SOI.
Therefore, large SHE and OHE due to such an effective flux will be 
ubiquitous in multiorbital systems.

We comment that the $(d_{xz},d_{yz})$-model has been applied to calculate the
large AHE in Ru-oxides \cite{Miyazawa,Kontani06}. Consistent with the theory,
paramagnetic compound Ca$_{1.7}$Sr$_{0.3}$RuO$_{4}$ exhibits large AHE 
under the magnetic field \cite{AHE-CSRO}.
We find that the same mechanism, that is, the effective Aharonov-Bohm phase
due to $d$-orbital degrees of freedom, gives rise to the large SHE and AHE.

In summary, we have studied SHC and OHC in the $t_{2g}$-model. 
The SHC obtained is
about three times larger than the recently observed \textquotedblleft giant
SHC\textquotedblright\ in Pt \cite{kimura}, and it can be further increased by
controlling ${\gamma}_{3}/{\gamma}_{1}$.
The large SHE in our model originates
from the \textquotedblleft effective Aharonov-Bohm phase\textquotedblright%
\ induced by the atomic SOI and the interorbital hopping integral. Moreover,
huge OHC in this model will enable us to control the atomic $d$-orbital state
by applying an electric filed.
The present calculation predicts that \textquotedblleft giant SHE and
OHE\textquotedblright\ will be ubiquitous in $d$-electron systems.
Using the theoretical technique developed here, SHC in various
transition metals such as Pt had recently been investigated \cite{SHE}.

The authors acknowledge fruitful discussions with
Y. Tanaka, S. Onari and T. Nomura. This work was supported by the Next
Generation Super Computing Project, Nanoscience Program, MEXT, Japan and
Grant-in-Aid for the 21st Century COE "Frontiers of Computational Science".


\end{document}